\documentclass[twoside]{article}
\usepackage{amsmath,amsfonts,amssymb,enumerate}
\usepackage{qic,epsfig}
\usepackage{cite}

\newcommand{\ket}[1]{| #1 \rangle}
\newcommand{\bra}[1]{\langle #1 |}

\begin{document}
\setlength{\textheight}{8.0truein}    

 \runninghead{A Decomposition Form of the Werner State}
            {name}

\normalsize\textlineskip \thispagestyle{empty}
\setcounter{page}{1}

\vspace*{0.88truein}

\alphfootnote

\fpage{1}

\centerline{\bf A Decomposition Form of the Werner State}
\vspace*{0.035truein} \centerline{\footnotesize
 Ming-Chung Tsai, Po-Chung Chen, Wei-Chi Su  and Zheng-Yao Su\footnote{Email: zsu@nchc.gov.tw}\hspace{.15cm}}
\centerline{\footnotesize\it Department of Physics, National Tsing
Hua University, Hsinchu, Taiwan, R.O.C.}
\centerline{\footnotesize\it National Center for High-Performance
Computing, Hsinchu, Taiwan, R.O.C.} \centerline{\footnotesize\it
National Center for Theoretical Sciences, Hsinchu and Tainan,
Taiwan, R.O.C.}

\vspace*{0.21truein}

 \abstracts{In this report, a scheme different from the PT and
 Wootters concurrence is developed to acquire a criterion to
 investigate the bipartite separability of the Werner state.}{}{}

 \vspace*{10pt}  \vspace*{3pt} \vspace*{1pt}\textlineskip

 \section{Introduction}\label{secintro}
  Entanglement is a type of characteristic that cannot be described
  by the classic physics.
  It is also known as an essential resource to a quantum
  processor and a quantum computation~\cite{}.
  A problem of great importance in the field of quantum information science
  is to qualitatively analyze entanglement.
  That is, determine whether a state
  is {\em separable} or {\em entangled}.
  According to the article of Werner in 1989~\cite{Werner1989},
  a state of a bipartite system ${\cal H}^{AB}$ is separable if it can be written as a {\em convex} combination of pure product states
 \begin{align}\label{eqdefnsep}
  \rho^{AB}=\sum^N_{i=1}p_i\rho^A_i\otimes\rho^B_i,
 \end{align}
 here $p_i\geq 0$, $\sum^N_{i=1}p_i=1$ and $\rho^{\mathcal{I}}_i$
 being a pure state in a Hilbert space
 $\mathcal{H}^{\mathcal{I}}$ with $(\rho^{\mathcal{I}}_i)^2=\rho^{\mathcal{I}}_i$ for $\mathcal{I}=A,B$.
 Otherwise, this state is {\em entangled}.
 The focus of this article will be on analyzing  he bipartitie separability of a specific class
 of states, the Werner states.

 There are a number of applications, based on the symmetry of the Werner state, to entanglement purification~\cite{BBPSSW,Short},
 nonlocality~\cite{Werner1989,Hiro,MFF}, entanglement measures~\cite{SST,CAF}, etc.
 Besides, another interest problem is how to examine the physical meaning
 of the critical point between the separable and entangled for the Werner states in a symmetry-qubit system.
 For this reason, an enormous number of research works have been realized
 to find out the decomposition formulation for the separable Werner states in a $2\times 2$ system~\cite{LS,Wootters,AzuBan,UKB}.
 Unanyan, etc.~\cite{UKB} proposes a decomposition formulation covering a part of the separable Werner states
 in a $3\times 3$ system.
 For a separable Werner state in a system of a higher dimension,
 constructing its decomposition formulation in terms of product states is a difficult task so far.
 In this study, we design a process to construct a form to represent
 an arbitrary separable Werner state in a $d\times d$ system.
 More specifically, this decomposition form will be demonstrated
 in the case that the dimension $d$ equals to a power of $2$.
 This decomposition is different from the decompositions introduced in~\cite{LS,Wootters,AzuBan,UKB}.
 Because of this convex combination formulation of separable Werner
 states were discussed in spinor basis, we keep our mind on the
 dimension Werner states.

 This article is organized as follows.
 In Sec.~\ref{secWernerSt}, we review the properties of Werner states and, for our purpose,
 we expand a Werner state in terms of the spinor representation.
 Besides, the separability of Werner states have been examined
 via Peres-Horodecki criterion~\cite{Peres,MPRHoros}.
 In Sec.~\ref{secDform}, the procedure to obtain the decomposition form of
 a Werner state will be demonstrated.
 Besides, the   and dimension Werner states were decomposed in this section.
 A novel and simple procedure to write down the convex combination
 formulation of product states for separable Werner states was inducted.
 In Sec.~\ref{secconclusion}, we give some respects for this decomposition formulation.

 \section{Werner State}\label{secWernerSt}
 A Werner state $\rho_W$ in a bipartite system ${\cal H}^{AB}$,
 ${\rm dim}{\cal H}^A={\rm dim}{\cal H}^B=d$,
 in general is defined as an invariant state under all bipartite local unitary transformations of the form
 $U\otimes U\in SU(d^2)$,
 that is,
 \begin{align}\label{eqWernerUU}
  \rho_W=(U\otimes U)\rho_W (U\otimes U)^\dag.
 \end{align}
 There are various kinds of explicit forms to represent a Werner state
 and the most popular one is
 \begin{align}\label{eqdefnWerner}
  \rho_W=\frac{1}{d^3-d}\{(d-f)I_A\otimes I_B+(df-1)P\},
 \end{align}
  where $I_A\otimes I_B$ is the identity in ${\cal H}^{AB}$
  and $P$ denotes the {\em flip} operator taking the form
 \begin{align}\label{eqflipOp}
  P=\sum^{d}_{i,j=1}\ket{i}\bra{j}\otimes\ket{i}\bra{j}.
 \end{align}
 Note that the Werner state $\rho_W$ is a one-parameter operator
 characterized by a real number $f={\rm Tr}_{AB}\{\rho_W\cdot P\}$.

  To obtain the required decomposition, we consider
  a Werner state $\rho_W$ in ${\cal H}^{AB}$ with $d=2^p$, $p\in \mathbb{N}$.
  In this case, the state can be written in the {\em spinor representation}.
  In terms of this representation, the Werner state $\rho_W$ in
  ${\cal H}_{2^p\times 2^p}$ is written as
 \begin{align}\label{eqWerner2p}
  \rho_W=\frac{1}{2^{3p}-2^p} \{ (2^p-f)I_{2^p\times 2^p} +\frac{2^pf-1}{2^p}
  \sum^{3}_{i_1,i_2,\cdots,i_p=0}\sigma_{i_1i_2\cdots i_p,i_1i_2\cdots i_p}\},
 \end{align}
  where the operator $\sigma_{i_1i_2\cdots i_p,i_1i_2\cdots i_p}$
  is defined in~\cite{SuTele,Su,SuTsai1,SuTsai2,SuTsai3,SuTsai4}.

 Each spinors of a Werner states in a $2^p\times 2^p$ system are commuting.
 Thus, there are linear relation $\vec{\lambda}=H^{\otimes p}M^{\otimes p}\vec{a}$
 between the eigenvalues of Werner states and coefficients of spinors.
 Here the two matrices $H$ and $M$ are $4\times 4$ matrices taking
 the forms
 \begin{align}\label{eqmatH&M}
  H=
  \begin{pmatrix}
   1&1&1&1\\
   1&-1&1&-1\\
   1&1&-1&-1\\
   1&-1&-1&1
  \end{pmatrix}
  \text{ and }
  M=
  \begin{pmatrix}
   1&0&0&0\\
   0&1&0&0\\
   0&0&1&0\\
   0&0&0&-1
  \end{pmatrix},
 \end{align}
 and $\vec{a}=(a_1,a_2\cdots,a_{2^p})^T$ denotes the vector of the coefficients of the
 spinors with $a_1=\frac{1}{2^{2p}}$ and
 $a_r=\frac{2^pf-1}{2^{4p}-2^{2p}}$ for $r>1$.
 The eigenvalues for the case $p=1$ are of the form
 \begin{align}\label{eignvalp=1}
  \lambda_1=\frac{1}{2^2}\{ 1-\frac{3(2f-1)}{2^2-1} \} \text{ and }
  \lambda_r=\frac{1}{2^2}\{ 1+\frac{(2f-1)}{2^2-1} \},
 \end{align}
 $2\leq r\leq 4$.
 While the eigenvalues for the case $p=2$ are
 \begin{align}\label{eignvalp=2}
  \lambda_s=\frac{1}{2^4}\{ 1-\frac{15(2^2f-1)}{2^4-1} \} \text{ and }
  \lambda_t=\frac{1}{2^4}\{ 1+\frac{3(2^2f-1)}{2^4-1} \},
 \end{align}
  here $1\leq s\leq 6$ and $7\leq t\leq 16$.
 Generally, the eigenvalues for the case $p\in\mathbb{N}$ take the
 form
 \begin{align}\label{eignvalp>2}
  \lambda_s=\frac{1}{2^{2p}}\{ 1-\frac{(2^p+1)(2^pf-1)}{2^{2p}-1} \} \text{ and }
  \lambda_t=\frac{1}{2^4}\{ 1+\frac{3(2^2f-1)}{2^4-1} \},
 \end{align}
  here $1\leq s\leq 2^{p+1}-2$ and $2^{p+1}-1\leq t\leq 2^{2p}$.
 The requisite of $\rho_W$ to be positive has the parameter $f$ in
 the range $-1\leq f\leq 1$.

 The Peres-Horodecki criterion or positive partial transpose (PPT) criterion~\cite{Peres,MPRHoros}
 gives a necessary condition to determine the separability of a bipartite density operator $\rho^{AB}$.
 That is, any separable state is still positive under the
 operation of partial transposition on party $A$ or $B$.
 According to this criterion, one derives the partial transpose of
 the Werner state~\cite{SuTele,Su,SuTsai1,SuTsai2,SuTsai3,SuTsai4}
 \begin{align}\label{eqPTWerner2p}
  \rho_W=\frac{1}{2^{3p}-2^p} \{ (2^p-f)I_{2^p\times 2^p} +\frac{2^pf-1}{2^p}
  \sum^{3}_{i_1,i_2,\cdots,i_p=0}(-1)^{\mu(i_1i_2\cdots i_p)}\sigma_{i_1i_2\cdots i_p,i_1i_2\cdots  i_p}\}.
 \end{align}
 When $p=1$, the eigenvalues of the partial transpose
 $\rho^{T_B}_W$ are calculated as
 \begin{align}\label{PTeignvalp=1}
  \lambda_1=\frac{1}{2^2}\{ 1+\frac{3(2f-1)}{2^2-1} \} \text{ and }
  \lambda_r=\frac{1}{2^2}\{ 1-\frac{(2f-1)}{2^2-1} \},
 \end{align}
 $2\leq r\leq 4$.
 As $p=2$, the eigenvalues of $\rho^{T_B}_W$ are
 \begin{align}\label{PTeignvalp=2}
  \lambda_s=\frac{1}{2^4}\{ 1+\frac{15(2^2f-1)}{2^4-1} \} \text{ and }
  \lambda_t=\frac{1}{2^4}\{ 1-\frac{3(2^2f-1)}{2^4-1} \},
 \end{align}
  here $1\leq s\leq 6$ and $7\leq t\leq 16$.
 The general form of the eigenvalues of $\rho^{T_B}_W$ are
 \begin{align}\label{eignvalp>2}
  \lambda_s=\frac{1}{2^{2p}}\{ 1+\frac{(2^p+1)(2^pf-1)}{2^{2p}-1} \} \text{ and }
  \lambda_t=\frac{1}{2^4}\{ 1-\frac{3(2^2f-1)}{2^4-1} \},
 \end{align}
  here $1\leq s\leq 2^{p+1}-2$ and $2^{p+1}-1\leq t\leq 2^{2p}$.
 As long as the Werner state $\rho_W$ is separable, the partial
 transpose $\rho^{T_B}_W$ is positive and the condition $0\leq f\leq 1$ is
 acquired.
 The condition $0\leq f\leq 1$ is the necessary condition for the
 separability of $\rho_W$ and in the next section, through the
 decomposition form, it is also the sufficient condition to
 determine the separability of $\rho_W$.

 \section{Decomposition Form}\label{secDform}
  In this section, the decomposition forms for the Werner states
  in a $2\times 2$ system will be first derived.
  Then, one decompose the Werner state in a $4\times 4$ system.
  The ns conditions for the bipartite separability will be
  discussed in both cases.
  Lastly, the decomposition form will be extended to the case of
  the Werner state in a $2^p\times 2^p$ system.

  Firstly, we demonstrate the decomposition form of a separable
  Werner state in a bipartite system ${\cal H}^{AB}$
  with ${\rm dim}{\cal H}^A={\cal H}^B=2$.
  Based on Eq.~\ref{eqWerner2p}, the Werner state, in terms of the spinor representation,
  is written as
 \begin{align}\label{eqWerner2by2}
  \rho_W=\frac{1}{2^3-2}\{ \frac{2^2-1}{2}I_2\otimes I_2+\frac{2f-1}{2}\sum^3_{i=1}\sigma_{i,i} \}.
 \end{align}
  One can decompose this state into the following form
 \begin{align}\label{eqWernerSepForm2by2}
  \rho_W=\frac{1}{6}\sum^3_{i=1}\sum_{\epsilon\in{Z_2}}\rho^A_{i,\epsilon}\otimes\rho^B_{i,\epsilon}
 \end{align}
  with the components
 \begin{align}\label{eqWernerSepFormAB2by2}
  \rho^A_{i,\epsilon}=\frac{1}{2}(I_2+(-1)^\epsilon\sqrt{|2f-1|}\sigma_i)
  \text{ and }
  \rho^B_{i,\epsilon}=\frac{1}{2}(I_2+(-1)^\epsilon\sqrt{|2f-1|}\sigma_i).
 \end{align}
  In this form, the state $\rho_W$ is separable iff all the
  components $\rho^A_{i,\epsilon}$ and $\rho^B_{i,\epsilon}$
  are positive operators.
  The implication is obvious that $\rho^A_{i,\epsilon}$ and $\rho^B_{i,\epsilon}$
  are positive if $\rho_W$ is separable.
  On the other implication, both the eigenvalues of the components $\rho^A_{i,\epsilon}$
  and $\rho^B_{i,\epsilon}$ take the form
 \begin{align}\label{eqWernerEValueB2by2}
  \lambda_{i,\epsilon}=\frac{1}{2}\{1+(-1)^\epsilon\sqrt{|2f-1|}\}
 \end{align}
  for $\epsilon\in{Z_2}$.
  The eigenvalues $\lambda_{i,\epsilon}$
  are thus positive as the parameter $f$ is in the range
  $0\leq f\leq 1$.
  Thus, the components $\rho^A_{i,\epsilon}$ and $\rho^B_{i,\epsilon}$
  are positive and $\rho_W$ is separable when $0\leq f\leq 1$.
  This exactly coincides with the result of the PPT criterion.
  Yet, the decomposition form is given to show the sufficient
  condition for the bipartite separability of the Werner state.

  Then, consider the case of higher dimension for ${\rm dim}{\cal H}^A={\cal H}^B=4$.
  By Eq.~\ref{eqWerner2p}, the Werner state in the spinor representation for $p=2$ is written as
 \begin{align}\label{eqWerner4by4}
  \rho_W=\frac{1}{2^6-2^2}\{ \frac{2^{4}-1}{2^2}I_4\otimes I_4+\frac{2^2f-1}{2^2}\sum^3_{i,j=0,(i,j)\neq(0,0)}\sigma_{ij,ij} \}.
 \end{align}
  This state cane be decomposed into the separable form
 \begin{align}\label{eqWernerSepForm4by4}
  \rho_W=\frac{1}{20}\sum_{\eta\in{Z_2}}\sum_{\epsilon\in{Z^2_2}}\sum^3_{i=1}\rho^A_{i,\epsilon\eta}\otimes\rho^B_{i,\epsilon\eta}
 \end{align}
  with the components
 \begin{align}\label{eqWernerSepFormAB4by4}
  &\rho^A_{i,\epsilon\eta}=\frac{1}{4}\{I_4+\sqrt{\frac{2^2f-1}{2^2-1}}
  [(-1)^{01\cdot\epsilon}\sigma_{i\eta}+(-1)^{10\cdot\epsilon}\sigma_{\eta i'}+(-1)^{11\cdot\epsilon}\sigma_{ii'}]\}
  \text{ and }\notag\\
  &\rho^B_{i,\epsilon\eta}=\frac{1}{4}\{I_4+\sqrt{\frac{2^2f-1}{2^2-1}}
  [(-1)^{01\cdot\epsilon}\sigma_{i\eta}+(-1)^{10\cdot\epsilon}\sigma_{\eta i'}+(-1)^{11\cdot\epsilon}\sigma_{ii'}]\},
 \end{align}
  here $i'=4-i$ and $\zeta\cdot\epsilon$ denoting the inner
  product of the two strings $\zeta$ and $\epsilon\in{Z^2_2}$.
  Similarly, the state $\rho_W$ is separable iff the operators
  $\rho^A_{i,\epsilon\eta}$ and $\rho^B_{i,\epsilon\eta}$
  are positive.
  A separable Werner state of course leads to positive operators
  $\rho^A_{i,\epsilon\eta}$ and $\rho^B_{i,\epsilon\eta}$.
  While, on the other hand,
  the eigenvalues of the components $\rho^A_{i,\epsilon\eta}$ and $\rho^B_{i,\epsilon\eta}$
  are of the forms
 \begin{align}\label{eqWernerEValueB4by4}
  \lambda_{i,\epsilon\eta}=\frac{1}{4}\{1+\sqrt{\frac{2^2f-1}{2^2-1}}
  [(-1)^{01\cdot\epsilon}+(-1)^{10\cdot\epsilon}+(-1)^{11\cdot\epsilon}]\}.
 \end{align}
  This has the component $\rho^A_{i,\epsilon\eta}$ and $\rho^B_{i,\epsilon\eta}$
  are positive when $\frac{1}{2}\leq f\leq 1$
  and then $\rho_W$ is separable.

  The decomposition form can be extended to the Werner state
  of the case ${\rm dim}{\cal H}^A={\rm dim}{\cal H}^B=2^p$.
  The form is written as
 \begin{align}\label{eqWernerSepForm2pby2p}
  \rho_W=\frac{1}{2^{3p}-2^{2p}}
  \sum^3_{i_1,\cdots,i_p=0}\frac{1}{2}
  \{\rho^A_{i,+}\otimes\rho^B_{i,+}+\rho^A_{i,-}\otimes\rho^B_{i,-}\}
 \end{align}
 with the component operators
 \begin{align}\label{eqWernerSepFormAB2pby2p}
  &\rho^A_{i,\pm}=\frac{1}{2^p}\{I_{2^p}\pm\sqrt{|2^pf-1|}\sigma_{i_1\cdots i_p}\}
  \text{ and }\notag\\
  &\rho^B_{i,\pm}=\frac{1}{2^p}\{I_{2^p}\pm\sqrt{|2^pf-1|}\sigma_{i_1\cdots  i_p}\}.
 \end{align}
  Likewise, the Werner state $\rho_W$
  is separable iff the components
  $\rho^A_{i,\pm}$ and $\rho^B_{i,\pm}$
  are positive operators.

  As a brief summary, the procedure of decomposing any separable Werner state
  can be itemized in the following steps.
  \begin{enumerate}
  \item Expand the Werner states in terms of the spinor representation.
  \item Distribute the weighted of identity operator to each spinor.
  \item Decompose each component containing the identity operator and spinors.
  \item Accomplish the decomposition of Werner states as $2^pf-1\leq 0$.
  \item Collect the spinors whose local generators in subspace and are commutable.
  \item Distribute the weighted of identity operator to each spinor.
  \item Decompose each component containing the identity operator and spinors.
  \item Accomplish the decomposition of Werner states as $2^pf-1\geq 0$.
  \end{enumerate}
  Notable, applying this procedure to the Werner state in any $2^p\times 2^p$ system.
  Form 1st step to 4th step, the decomposition formulation for the separable Werner states can be obtained
  as $0\leq f\leq \frac{1}{2^{p-1}}$.
  From 5th step to 8th step, the number of spinor whose local generators in ${\cal H}^A$ and ${\cal H}^B$
  are commuting is $2^p-1$.
  One could obtain $2^p+1$ components that contain commutative spinors.
  For this reason, the identity operator $I$ can be divided to $2^p+1$  and assigned to each component.
  The decomposition formulation for the separable Werner state in a $2^p\times 2^p$ system is thus obtained
  as $\frac{1}{2^p}\leq f\leq 1$.
  Therefore, this approach could decomposed all separable Werner state in a $2^p\times 2^p$ system.

 \section{Conclusion}\label{secconclusion}
  In this article, a novel and simple approach to write down the decomposition
  formulation for separable  dimension Werner states.
  The main idea of this approach include:
  1. expanding the Werner states in spinor basis;
  2. dividing the spinor into some components;
  3. distributing the weighted of identity operator to each component;
  4. decomposing the each component.
  The sufficient condition of separable Werner states could be established by the positive eigenvalues of local operators.
  This decomposition approach is validating for separable   dimension Werner states in  .
  It is still a challenge problem to write down a decomposition formula of all separable Werner states in any dimension.
  However, under the appropriate choices of basis,
  the correct decomposition formulation of Werner states in arbitrary dimension would probably obtain.

\nonumsection{References}

 \end{document}